\documentclass[10pt]{iopart}

\usepackage{mathrsfs}
\usepackage{amssymb}
 \usepackage{graphics}
 \usepackage{graphicx}
 \usepackage{epsfig}
\usepackage{iopams}  
\usepackage{citesort}

 \newtheorem{definition}{Definition}[section]
 \newtheorem{lemma}{Lemma}[section]
 
 \newtheorem{theorem}{Theorem}[section]
 \newtheorem{remark}{Remark}[section]

 \newtheorem{corollary}{Corollary}[section]

\begin{document}

\title[Degenerated ground-states in a class of 1D Ising-like models]{Degenerated ground-states in a class of 1D Ising-like models: a characterization by symbolic dynamics}
\author{L. A. Corona}

\address{Instituto de Investigaci\'on en Ciencias B\'asicas y Aplicadas. Universidad Aut\'onoma del Estado de Morelos, Avenida Universidad 1001, Cuernavaca Morelos, C.P. 62209, Mexico.}

\author{R. Salgado-Garc\'ia}
\ead{raulsg@uaem.mx}
\address{Centro de Investigaci\'on en Ciencias. Universidad Aut\'onoma del Estado de Morelos, Avenida Universidad 1001, Cuernavaca Morelos, C.P. 62209, Mexico.}

\date{\today}                                          % Activate to display a given date or no date

%
%\vspace{10pt}
%\begin{indented}
%\item[]February 2014
%\end{indented}
%

\begin{abstract}

In this note we study a class of one-dimensional Ising chain having a highly degenerated set of ground-state configurations. The model consists  of spin chain having infinite-range pair interactions with a given structure. We show that the set of ground-state configurations of such a model can be fully characterized by means of symbolic dynamics. Particularly we found that the set  ground-state configurations defines what in symbolic dynamics is called \emph{sofic shift space}. Finally we prove that this system has a non-vanishing residual entropy (the topological entropy of the shift space), which can be exactly calculated. 

\end{abstract}

% Uncomment for PACS numbers
%
% Uncomment for keywords
%\vspace{2pc}
%\noindent{\it Keywords}: XXXXXX, YYYYYYYY, ZZZZZZZZZ
%
% Uncomment for Submitted to journal title message
\submitto{\JPA}

%\date{\today}                                          % Activate to display a given date or no date

%
% Uncomment if a separate title page is required
\maketitle
% 
% For two-column output uncomment the next line and choose [10pt] rather than [12pt] in the \documentclass declaration
%\ioptwocol
%

%#############################################################
%#############################################################
%
\section{Introduction}
%
%#############################################################
%#############################################################

It has long been known that several models of spin systems might have highly degenerated set of ground-state configurations at a critical magnetic field~\cite{domb1960theory}. One of the earlier examples of the occurrence of such a phenomenon was found in one-dimensional spin chains with nearest-neighbor interactions (specifically, in spin chains with range-one interactions) at a critical magnetic field~\cite{brooks1951order,fisher1960lattice,domb1960theory}. The importance of this phenomenon is due to the fact that it violates the third law of thermodynamics in the sense that the entropy of the system must vanish at the zero absolute temperature. The occurrence of such a phenomenon, in spin systems particularly~\cite{slegers1988residual,hajdukovic1982ground,redner1981one}, led to some authors to reexamine such a basic principle~\cite{aizenman1981third,chow1987residual,wreszinski2009precise}. Nowadays it is accepted that a thermodynamical system can have a non-zero entropy at zero absolute temperature, which is referred to a the \emph{residual entropy}~\cite{wreszinski2009precise}. In fact, the residual entropy is related directly to the degeneracy of the set ground-state configurations of the system. 

Within the context of the thermodynamic formalism, it is known that a spin system on a regular lattice can be viewed as a symbolic dynamical system endowed with a certain function characterizing the interactions among spins. Within this setting, it is known that the set of ground-state configurations (called hereafter simply the \emph{groundstate}~\footnote{Do not confuse the concept of groundstate defined here with the one defined in the thermodynamical formalism. In the last context, a \emph{groundstate} refers to the limit of the Gibbs measure  as the temperature goes down to zero. Here, call groundstate to the set of all configurations that minimize the energy of the system, which contains the support of the zero-temperature limit of the Gibbs measure of the system (if such a limit does exists).}) has well defined mathematical properties. For example, it is known that a groundstate is actually a shift space, i.e., it is a subset of the set of all spin configuration which is invariant under the shift mapping. Moreover, it is known that spin systems having finite-range two-body interactions have in general a groundstate which turns out to be  a finite union of subshift of \emph{finite type}~\cite{garibaldi2012description,chazottes2011zero}. In general, spin systems might have groundstates which are not necessarily subshifts of finite type. For example, in Ref.~\cite{van1998nonperiodic} it has been proved that a discrete spin system with infinite-range four-body interaction has a non-periodic groundstate. Indeed the authors of Ref.~\cite{gardnert1989fractal} proved that such a groundstate is a Thue-Morse subshift whose residual entropy is zero. Indeed, one can build spin systems, with a rather artificial form the interactions among spins, whose groundstates can be actually any subshift space~\cite{chazottes2010zero}. This is done through a class of functions called \emph{Lipschitz functions} which to spin systems are not necessarily a physically realistic type of interactions. 

In this work we provide an example of a spin system with infinite-range pair interactions having a groundstate which turns out to be a \emph{strictly sofic} subshift. The system that we introduce here and its groundstate has some interesting properties. Particularly, our model has two-body (arbitrarily fast decaying) interactions which can be considered physically realistic. The ground-state configurations can be fully characterized  through symbolic dynamics techniques, and the residual entropy can be exactly determined. Recall that a sofic subshift is defined as a factor of a subshift of finite type~\cite{weiss1973subshifts}. Thus any subshift of finite type is sofic. A strictly sofic subshift is therefore a sofic subshift that is not of finite type. The main difference between strictly sofic subshift and a subshift of finite type is that the later can be characterized by a finite set of forbidden \emph{words}, while the former needs an infinite set of forbidden words to be characterized~\cite{lind1995introduction}. 

This work is organized as follows. In section~\ref{sec:setting} we state the model, we give some basic definitions on symbolic dynamics and state the notation that we will use throughout this work. In section~\ref{sec:results} we sate the main results of this work and finally in section~\ref{sec:proofs} we give the proof of our results.

%#############################################################
%#############################################################
%
\section{Setting and generalities}
\label{sec:setting}
%
%#############################################################
%#############################################################

%#############################################################
\subsection{Preliminary concepts}
%#############################################################

In statistical mechanics an interaction potential among spins on a one-dimensional lattice is a family of shift-invariant functions $\Phi = (\Phi_\Lambda)_{\Lambda\subset \mathbb{Z}}$ indexed by finite subsets $\Lambda$ of $\mathbb{Z}$~\cite{ruelle1978thermodynamic}. On the other hand, within the thermodynamical formalism, is a \emph{potential} which describes the interactions on a given spin system. A potential is a function $\psi : \Sigma \to \mathbb{R}$ on the set of all configurations of infinite spin chains $\Sigma := \{+,-\}^\mathbb{Z}$ to the real line. Of course, these two concept are related each other~\cite{ruelle1978thermodynamic}. Given a family of interactions on a one-dimensional spin system $\Phi$, the potential $\psi$ is obtained as follows,
\[
\psi = \sum_{\Lambda \ni 0} \frac{\Psi_\Lambda}{\#\Lambda}
\]
where $\#\Lambda $ stands for the cardinality of $\Lambda$. In this work we will adopt the last point of view, i.e.,  our spin system will be determined by a potential function in the sense of the thermodynamical formalism. 

Let us denote by $\boldsymbol{\sigma} = (\dots \sigma_{-1}\sigma_{0}\sigma_{1}\dots)$ an element of $\Sigma$. Thus, $\boldsymbol{\sigma} \in \Sigma$ represents the infinitely long spin chain and we can think of every coordinate of $\boldsymbol{\sigma}$ as a spin variable. Each spin $\sigma_i$ interacts with the rest of spins via a set of pair interactions. An infinite-range pair interaction potential on this spin chain can be defined as a function on the full shift to the real line, $\psi : \Sigma \to \mathbb{R} $, which can be written with certain generality as
\begin{equation}
\label{eq:potential-physical}
\psi(\boldsymbol{\sigma} ) := H\sigma_0 + \sum_{j=1}^\infty K(j)\sigma_0 \sigma_j.
\end{equation}
Here $K(j)$ is the coupling constant for the interaction between the $0$th spin and the $j$th spin on the chain. The first term in the above equation, $H\sigma_0$ represents the interaction of the zeroth spin with an external magnetic field. We say that the interaction $\psi$ is summable if 
\begin{equation}
|| \psi || := |H| + \sum_{j=1}^\infty |K(j)| < \infty.
\end{equation}

Now let us introduce some basic concepts and notation on symbolic dynamics. Let $\mathcal{A} $ a finite set, to which we will refer to as \emph{alphabet}. We denote by $\mathcal{A}^\mathbb{Z}$ the set of all infinite symbolic sequences made up from elements of the alphabet  $\mathcal{A}$. An infinite symbolic sequence $\mathbf{x} \in \mathcal{A}^\mathbb{Z}$  is also written as $\mathbf{x} = \dots x_{-1}x_0 x_1\dots$, where each $x_j$ is an element from the alphabet $\mathcal{A}$. Each $x_j$ in $\mathbf{x}$ is referred to as a \emph{coordinate} of $\mathbf{x}$ or as the \emph{$j$th coordinate} of $\mathbf{x}$ if we would like to emphasize its location along $\mathbf{x}$. The $j$th coordinate of the symbolic sequence $\mathbf{x}$ is alternatively written as $ (\mathbf{x})_j:=x_j$. We call $\mathcal{A}^n$ the set of all finite symbolic sequences made up of $n$ symbols. A finite string $\mathbf{a} = a_0a_1\dots a_{n-1}$ in $\mathcal{A}^n$ will also be referred to as a \emph{word} or as a \emph{block} of size $|\mathbf{a}|:= n$. If $\mathbf{x} \in \mathcal{A}^\mathbb{Z}$ is a symbolic sequence we denote the finite string $x_jx_{j+1}\dots x_{j+n-1}$ as $\mathbf{x}_j^{j+n-1}$. Clearly $\mathbf{x}_j^{j+n-1}$ is an element of $\mathcal{A}^n$. We say that a word $\mathbf{a}\in\mathcal{A}^n$ \emph{occurs} in $\mathbf{x} \in \mathcal{A}^\mathbb{Z}$ if there is a $j\in\mathbb{Z}$ such that $\mathbf{x}_j^{j+n-1} = \mathbf{a}$. We also say that  $\mathbf{x}$ has as \emph{prefix} $\mathbf{a}$ if $\mathbf{a}$ occurs in $\mathbf{x}$ for $j=0$. The \emph{concatenation} of two words $\mathbf{a} \in \mathcal{A}^n $ and $\mathbf{b} \in \mathcal{A}^m $ is an operation, denoted as a \emph{multiplication} between words, that gives a new word $\mathbf{c} := \mathbf{a} \mathbf{b} $  in the set $\mathcal{A}^{n+m}$ for any $n,m\in\mathbb{N}$. It is clear that this ``multiplication'' of words is not commutative, since in general $\mathbf{a} \mathbf{b}  \not= \mathbf{b} \mathbf{a} $. The \emph{exponentiation} of words should be understood as a concatenation of a word with itself, if $\mathbf{a} \in \mathcal{A}^n $  then $\mathbf{a}^m = \mathbf{a}\mathbf{a}\mathbf{a}\cdots\mathbf{a} $ is a word in the set $ \mathcal{A}^{nm}$. The exponentiation of a word to the $0th$ power, $\mathbf{a}^0$, will be defined as the empty word by convenience. 

Given a word $\mathbf{a} \in \mathcal{A}^n$ a word of size $n$ we define a \emph{cylinder set} (or simply a \emph{cylinder}) $[\mathbf{a}]$ as the subset of $\mathcal{A}^\mathbb{Z}$ containing all the symbolic sequences having as prefix the word $\mathbf{a}$, i.e.,
\[
[\mathbf{a}] := \{ \mathbf{x}\in \mathcal{A}^\mathbb{Z}\, : \, \mathbf{x}_0^{n-1} = \mathbf{a} \}.
\]
The collection of all cylinder sets forms a basis for a topology for $\mathcal{A}^\mathbb{Z}$ which makes $\mathcal{A}^\mathbb{Z}$ a compact space. Actually, the cylinder sets are closed an open set within this topology. 

The shift mapping $T:\mathcal{A}^\mathbb{Z} \to \mathcal{A}^\mathbb{Z}$ is a function that shift the sequence to the left with respect to the index. In other words, if $\mathbf{x} \in \mathcal{A}^\mathbb{Z}$ is a point on the full shift, then $T(\mathbf{x})$ is another point on $\mathcal{A}^\mathbb{Z} $ such that $T(\mathbf{x})_j = x_{j+1}$ for all $j\in\mathbb{Z}$.

If $Y\subset \mathcal{A}^\mathbb{Z} $, we say that $(Y,T)$ is a shift space if it is invariant and closed set under the shift mapping $T$. The full shift $\mathcal{A}^\mathbb{Z}$ itself is a shift space and it is usually referred to as the \emph{full shift}. A shift space (or subshift) $Y$ can also be defined through a collection of \emph{forbidden words} in the sequences contained in $Y$. We denote the set of forbidden words for $Y$ as $F(Y)$. In particular, for the full shift $\mathcal{A}^\mathbb{Z}$, the set of forbidden words is the empty set. Those shift spaces having a finite collection of forbidden words are said to be of \emph{shifts of finite type}. The class of shifts spaces that we will consider here are known as \emph{sofic shifts}. A shift space $Y$ is said to be \emph{sofic} if there is a shift of finite type $\tilde Y$ and an onto mapping $\alpha : \tilde Y \to Y$ such that the $0$th coordinate of $\alpha (\mathbf{a})$ depends on a finite number of coordinates of $\mathbf{a}$, for all $\mathbf{a} \in \tilde Y$, and $\alpha\circ \tilde T = T \circ \alpha$ where $\tilde T$ and $T$ stands for the shift actions on $\tilde Y$ an  $Y$ respectively. Then, a sofic shifts is \emph{factor} of a shift of finite type. It can be show that sofic shifts can also be characterized by an infinite set of forbidden words and more specifically through the concept of follower sets (see for example Ref.~\cite{lind1995introduction}).

Finally, a different way to characterize a shift space is through the concept of \emph{language}. Let $Y$ be a subshift space, then, the \emph{language} $\mathcal{B}(Y)$ for $Y$ is the collection of all the words occurring in any element of $Y$. Particularly we denote by  $\mathcal{B}_n(Y)$ the set of all the words of size $n$ occurring in any element of $Y$. In this way we can say that $\mathcal{B}(Y) = \cup_{n\in\mathbb{N}} \mathcal{B}_n(Y) $.

%#############################################################
\subsection{The Ising-like spin model}
%#############################################################

Let us consider the full shift $\Sigma := \mathcal{A}^\mathbb{Z} $ with the alphabet of all possible spin configurations, $\mathcal{A} = \{+,-\}$. A \emph{coboundary} is a function $f : \Sigma \to \mathbb{R}$ such that there is a function $g : \Sigma \to \mathbb{R}$ for which, $f = g - g\circ T$. It is know that two potentials which differ in a coboundary are physically equivalent in the sense that they define the same Gibbs measure and have the same set of ground-state configurations~\cite{bousch2001condition,sarig1999thermodynamic}. It is said that two potentials that differs in a coboundary are \emph{cohomologous}.  

We can profit from the concept of coboundary to simplify our problem by means of a potential on a different shift space with an equivalent groundstate. First, recall that the spin variables $\sigma_i$ along the infinite spin chain can take the values $+1$ and $-1$. Let us define new variables $x_i$ taking values in the set $\{0,1\}$ through the transformation
\begin{equation}
\sigma_i = f(x_i) = 2x_i - 1, \qquad \forall \ i \in \mathbb{Z}.
\end{equation}
It is clear that the symbol `0' corresponds to the state $-1$ and the symbol `1' corresponds to the state $+1$ of the corresponding spin variable. The above transformation on a single symbol (or a single coordinate) can be extended coordinatewise to the whole spin chain. This extended transformation, which we will denote by $F : \{0,1\}^\mathbb{Z} \to \{-1,+1\}^\mathbb{Z} $, establishes an isomorphism between $\Sigma $ and $\{0,1\}^\mathbb{Z} $.  If we introduce this ``change of variable'' in the potential~(\ref{eq:potential-physical}) we obtain a new potential $\phi : \{0,1\}^\mathbb{Z}  \to \mathbb{R}$ on the full shift $X :=\{0,1\}^\mathbb{Z}$. Thus, given a symbolic sequence $\mathbf{x} = \cdots x_{-1}x_{0}x_{1} \cdots \in X$, the potential $\phi(\mathbf{x})$ is defined as, 
\begin{equation}
\tilde  \phi(\mathbf{x}) := \psi \circ F (\boldsymbol{\sigma}),
\end{equation}
which is explicitly given by,
\begin{eqnarray}
\tilde \phi(\mathbf{x}) &=& H f(x_0) + \sum_{j\in \mathbb{Z}}  K(j)f(x_0) f(x_j),
\nonumber
\\
 &=& H (2x_0 - 1) + \sum_{j\in \mathbb{Z}}  K(j)(2x_0 - 1)(2x_i - 1),
\nonumber
\\
 &=& 2 H x_0 - H+ \sum_{j\in \mathbb{Z}}  K(j)\left( 4x_0 x_j - 2(x_0 + x_j) + 1\right).
\end{eqnarray}
Performing some additional calculations we obtain,
\[
\tilde \phi(\mathbf{x}) =2 H x_0 - H + \sum_{j\in \mathbb{Z}}  4K(j) x_0 x_j - 2 K_0 x_0  -2 \sum_{j\in \mathbf{Z}}  K(j)x_j + ||\psi||,
\]
or equivalently
\[
\tilde \phi(\mathbf{x}) =2 (H -K_0) x_0 + \sum_{j\in \mathbb{Z}}  4K(j) x_0 x_j - 2 \sum_{j\in \mathbb{Z}}  K(j)x_j + (K_0 - H),
\]
where $K_0$ is a constant defined as
\[
K_0 := \sum_{j=1}^\infty K(j).
\]
Now, let us define a new potential the potential  $\phi$ which will be physically equivalent to the potential  $\tilde \phi$. These two potential should be cohomologous or, in other words, they should differ in a coboundary, i.e., 
\begin{equation}
\label{eq:first-phi}
\phi = \tilde \phi + \xi - \xi\circ T
\end{equation}
for some well-behaved real-valued function $\xi : X \to \mathbb{R}$. Let us chose the function $\xi $ as, 
\[
\xi (\mathbf{x}) := 2 \sum_{ j =0}^\infty L(j) x_j, 
\]
where
\[
L(j) := -\sum_{n=j+1}^\infty K(n).
\]
Then, notice that this function defines the coboundary, 
\begin{eqnarray}
\xi(\mathbf{x}) - \xi\circ T(\mathbf{x}) &=& 2L(0) x_0 + \sum_{j=1}^\infty 2 \left( L(j)-L(j-1)\right)x_j,
\nonumber
\\
&=& -2K_0 x_0 + \sum_{j=1}^\infty 2 K(j) x_j.
\end{eqnarray}
where we have identified $L(0)$ with the constant $K_0$. Thus, if we add this coboundary to the potential $\tilde \phi$ we obtain,
\begin{equation}
\phi (\mathbf{x}) :=  2 (H - 2 K_0) x_0  + (K_0 - H) + \sum_{j = 1}^\infty  4K(j) x_0 x_j.
\end{equation}

If we  chose the magnetic field $H$ at the \emph{critical value} $H_c := 2 K_0$, we have that the potential become
\begin{equation}
\label{eq:phi-critical}
\phi_{\mathrm{c}} (\mathbf{x}) = -\frac{H_{\mathrm{c}}}{2} + \sum_{j = 1}^\infty  J(j) x_0 x_j, 
\end{equation}
where we have defined the coupling constant $J(j)$ as $J(j) := 4K(j)$. This result means that the original spin model at the critical magnetic field $H_c$ is physically equivalent to a symbolic chain, each lattice variable having one of the possible states `0' and `1', with interactions given by the potential~(\ref{eq:phi-critical}). 

The potential $\phi_{\mathrm{c}}$ is still too general to be treated. Indeed, we will restrict the coupling constant to a specific type defining a large class of potentials on $X$.  In this work we will consider the class of potential given by~(\ref{eq:phi-critical}) with coupling constants between lattice sites satisfying that,
\[
J(qj) = 0 \quad \forall \  j \in \mathbb{N}, \qquad \mbox{and} \qquad J(j) > 0 \quad \mbox{otherwise},
\]
for some positive $q \geq 2$.
It is clear that in order for the potential $\phi_{\mathrm{c}}$ to be well defined we need to verify the summability condition on $\phi$. Indeed, a direct calculation shows that the summability for $\psi$ (which states that $||\psi ||<\infty$) implies the summability for $\phi$. However, in order for the coboundary to be well defined, we also need to verify the summability condition for $\xi$. The last condition leads us to impose that the coupling constants $J(j)$ also satisfy, 
\[
\sum_{j=0}^\infty L(j) \leq \sum_{j=0}^\infty \sum_{k=j+1}^\infty |K(j)| = \frac{1}{4}\sum_{j=0}^\infty \sum_{k=j+1}^\infty |J(j)| < \infty.
\]
It is clear that the above condition automatically implies that $||\psi|| <\infty$, by which we will only need to assume the last one without loss of generality. We state this condition explicitly below.

\textbf{Condition 1}\textit{ We say that a potential $\phi_{\mathrm{c}}  : \{0,1\}^\mathbb{Z} \to \mathbb{R}$, defined as}
\[
\phi_{\mathrm{c}} (\mathbf{x}) = -\frac{H_{\mathrm{c}}}{2} + \sum_{j = 1}^\infty  J(j) x_0 x_j,
\]
\textit{satisfies the summability condition if}
\[
\sum_{j=0}^\infty \sum_{k=j+1}^\infty |J(j)| < \infty.
\]

Additionally, we should also emphasize that the constant term $H_{\mathrm{c}}/2$ in $\phi_{\mathrm{c}}$ can be neglected since any constant term added to a given potential, define the same Gibbs measure as the original potential and have the same set of ground-sate configurations. In this way, two potential which differs in a constant can be considered as physically equivalent.

%#############################################################
%#############################################################
%
\section{Main result}
\label{sec:results}
%
%#############################################################
%#############################################################

Along the rest of this work we will denote by $X$ the binary full shift, $X:= \{0,1\}^\mathbb{Z}$.  
Let $\phi : X \to \mathbb{R}$ be a potential on $X$.  Given $p\in \mathbb{N}$ and $\mathbf{x} \in X$ we define $S_p\phi (\mathbf{x})$ as the partial ergodic sum,
\[
S_p\phi(\mathbf{x}) := \sum_{j=0}^{p-1} \phi\circ T^j(\mathbf{\phi}).
\]
If $\phi(\mathbf{x})$ is interpreted as the interaction energy of the $0$th spin with all the other spins in the chain, then, the function $S_p\phi(\mathbf{x})$ can be interpreted as the total energy of a block of spins $x_0x_1\dots x_{p-1}$ of size $p$. In this way we can think of $S_p\phi$ as the equivalent to what in statistical physics is called \emph{Hamiltonian} of a block of spins of size $p$. In the same line of interpretation, we can say that $S_p\phi(\mathbf{x})/p$ corresponds to the mean energy per spin in the referred block of spins.  

Let us also call $\mathrm{Per}_p(X)\subset X$ to the set of all the periodic points $x\in X$, under $T$, of period $p$. This set is equivalent to the set of all infinite sequences which can be seen as a block of size $p$ infinitely repeated along the symbolic sequence,
\[
\mathrm{Per}_p:=\{\mathbf{x}\in X \,:\, T^p(\mathbf{x}) = \mathbf{x}\}
\]

We also define the \emph{minimal mean energy per spin}  $\phi_{\mathrm{min}}$ as~\cite{jenkinson2006ergodic,chazottes2011zero},
\[
\phi_{\mathrm{min}} := \inf \bigg\{ \min\bigg\{\frac{S_p\phi(\mathbf{x})}{p}\,:\,\mathbf{x} \in \mathrm{Per}_p(X) \bigg\} \,:\, n\in\mathbb{N} \bigg\}
\]

Now we proceed to define what we will call groundstate in the sense described above. 

\begin{definition}
Let  $\phi : X \to \mathbb{R}$ be a potential on the full shift $X$. The groundstate of the potential $\phi$ (also referred to as the $\phi$-minimizing subshift) $\underline{X}(\phi)$, is defined as
\[
\underline{X}(\phi) := \mathrm{clos}\bigg( \bigcup_{n \in\mathbb{N}} \{ \mathbf{x} \in \mathrm{Per}_p(X) \, :\, S_p\phi(\mathbf{x}) = p\phi_{\mathrm{min}}  \} \bigg).
\]

\end{definition}

It is clear from the above definition that the $\phi$-maximizing sets are $T$-invariant, and then, by the closure action, shift spaces. With this definition of groundstates we can state the following result.

\begin{theorem}

Let  $\phi_{\mathrm{c}} : X \to \mathbb{R}$ be a potential on the full shift $X$ defined as
\[
\phi_{\mathrm{c}} (\mathbf{x}) :=  \sum_{j=1}^\infty J_{\mathrm{c}}(j) x_0 x_j,
\]
satisfying Condition 1. Then, the groundstate for $\phi_{\mathrm{c}}$ is given by,
\[
Y^{(q)} := \mathrm{clos}\bigg( \bigcup_{n\in\mathbb{N}} {Y}_{n}^{(q)} \bigg),
\]
i.e., $\underline{X}\left(\phi_{\mathrm{c}}\right) = Y^{(q)}$. Here the family of sets ${Y}_{n}^{(q)}  $ are defined as,
\[
\fl
 {Y}_{n}^{(q)} := \{ \mathbf{x}\in \mathrm{Per}_n(X) \, :\, \mathbf{x}_j^{j+kq+l} \not = 10^{kq+l-1}1, \,\, \mbox{for} \,\, 1\leq l \leq q-1, \,\,  \forall \, k\in \mathbb{N}  \,\,\mbox{and}\,\, \forall \, j\in \mathbb{Z} \}.
\]
\label{teo1}
\end{theorem}

The above theorem states that the model we propose have a sofic groundstate, and moreover, it state the explicit form of the subshift by means of an infinite set of forbidden words. Actually, each set $Y^{(q)}_n$ can be seen as defined through a set of ``forbidden words''. The forbidden words turns out to be of the form $10^m1$ with $m$ not as a multiple of $q$ minus 1, i.e. $m = kq+l-1$ for all $1\leq l \leq q - 1$ and for all $k\in\mathbb{N}_0$. Thus, taking the closure of the union of all the sets $Y^{(q)}_n$ we obtain a shift space with a set forbidden words which turns out to be infinite. Actually, the sets of forbidden words for $Y^{(q)}$ can be written as
\[
F(Y^{(q)}_n) = \{ 10^{kq+ m-1}1 \, : \, \forall \, k\in \mathbb{N}_0, \forall\, 1\leq m \leq q-1\}.
\]
Moreover, we will see below that the set of forbidden words can be seen, as an extension of the above, as
\[
\tilde{F}(Y^{(q)}_n) = \{ 1\mathbf{a}1 \, : \, \forall \, \mathbf{a} \in\mathcal{A}^{kq+m-1},\, \forall \, k\in \mathbb{N}_0, \forall\, 1\leq m \leq q-1\}.
\]

Then, it becomes clear that the resulting groundstate is not of finite type because its set of forbidden words is not finite. However, this is not enough to say that the groundstate is sofic. Actually, to state the sofic property for $\underline{X}(\phi_{\mathrm{c}})$  we need to characterize explicitly the (finite)  family of follower sets.

\begin{theorem}

The groundstate $\underline{X}(\phi_{\mathrm{c}}) = Y^{(q)}$ is sofic and all its follower sets are given by 
\begin{eqnarray}
\qquad \quad \, \, \,
\mathcal{F}(0^q)&:=&\mathcal{B}(Y^{(q)}),
\nonumber
\\
\mathcal{F}(0^{m}10^{q-m-1})&=&\bigcup_{n\in\mathbb{N}} \mathcal{C}_{n,m}^{(q)},
\label{eq:followersets}
\end{eqnarray}
where the sets  $\mathcal{C}_{n,m}^{(q)}$ are defined as
\[
\fl
\mathcal{C}_{n,m}^{(q)} := \left\{ \mathbf{a} = a_0a_1a_2\dots a_{n-1} \in \mathcal{B}_n(Y^{(q)}) \, : \, a_i =0 \, \forall \,i \not= kq + m,\, \forall\, 0\leq k\leq \lfloor n/q \rfloor\right\}.
\]
\label{teo:t2}
\end{theorem}

The family of follower set  allows us to build up the corresponding right-resolving presentation for the sofic subshift  $\underline{X}(\phi_{\mathrm{c}})=Y^{(q)}_n$. Then, the corresponding graph associated to the subshift $Y^{(q)}$~\footnote{The graph associated to the subshift $Y^{(q)}$ means that the subshift $Y^{(q)}$  is a factor of the subshift of finite type that defines such a graph.} allows us to calculate explicitly the topological entropy of the system~\cite{lind1995introduction}. 

\begin{corollary}
\label{coro-1}
The residual entropy of the groundstate $\underline{X} (\phi_{\mathrm{c}}) $ is $ h_{ \mathrm{top} }(\underline{X}(\phi_{\mathrm{c}})) = \log(2)/q$.
\end{corollary}

%#############################################################
%#############################################################
%
\section{Proof of the main result}
\label{sec:proofs}
%
%#############################################################
%#############################################################

Before giving the proof of the main theorem, let us state two result which will allow us to understand better the structure of the subshift $Y^{(q)}$ as well as the groundstate for $\phi_{\mathrm{c}}$. 

\begin{lemma}
The minimal mean energy per spin for $\phi_{\mathrm{c}}$ is $\phi_{\mathrm{min}} = 0$.
\end{lemma}
%%%%%%%%%%%%%%%%%%%%%%%%%%%%%%%%%%%%%
\emph{Proof:} 

Notice that if we take a $\mathbf{x}\in \mathrm{Per}_n(X)$,  then $\phi_{\mathrm{min}} \leq S_n\phi_{\mathrm{c}} (\mathbf{x})/n$.  Take $\mathbf{x} = 0^\infty$. Then, it is clear that $\phi_{\mathrm{min}} \leq 0$. On the other hand it is clear that $\phi_{\mathrm{min}} \geq  \min\{\phi_{\mathrm{c}} (\mathbf{x}) \, :\, \mathbf{x}\in X\}$. However we have that $ \min\{\phi_{\mathrm{c}} \, :\, \mathbf{x}\in X\} = 0$. This implies the result.

\hfill\ensuremath{\square}
%%%%%%%%%%%%%%%%%%%%%%%%%%%%%%%%%%%%%

The above lema state that the mean energy per spin for the groundstate is zero. Therefore, all the  $n$-periodic points $\mathbf{x}$ belonging to the groundstate should satisfy that $S_p\phi_{\mathrm{c}}(\mathbf{x})=0$. This fact will be used below for the proof of the main theorem.

\begin{lemma}
If $n$ is not a multiple of $q$, then $Y^{(q)}_n=\{0^\infty\}$. 
\label{lemma-1}
\end{lemma}

%%%%%%%%%%%%%%%%%%%%%%%%%%%%%%%%%%%%%
\emph{Proof:} 
To prove this statement, first notice that $0^\infty \in Y^{(q)}$, since in $0^\infty$ does not occur any word of the type $10^{kq+l-1}1$. Thus, $\{0^\infty\} \subset Y^{(q)}$. On the other hand, let us assume that there is a periodic point $\mathbf{x} \in Y_n^{(q)} $ with period $n=k_0 q+m_0$, with $1\leq m_0 \leq q-1$, for which there is a $j_0\in \mathbb{Z}$ such that $x_{j_0}= 1$. By periodicity we have that $x_{j_0+k_0q} = 1$. Then notice that $\mathbf{x}$ contains a word of the form $1\mathbf{a}_01$ with $ |\mathbf{a}_0| = k_0q+m_0-1$ for $1\leq m_0 \leq q-1$, specifically, $\mathbf{x}_{j_0}^{j_0+k_0q+m_0} = 1\mathbf{a}_01$. But, by definition of $Y^{(q)}_n$, no word of the form $10^{kq+l-1}1$ occurs in $\mathbf{x}$ for $1\leq l\leq q-1$ and all $k\in\mathbb{N}_0$. This meas that $ \mathbf{a}_0 $ cannot be a word of the form $0^{k_0q+m_0-1}$ since it is ``forbidden'' for the elements of $Y^{(q)}_n$. Thus, there would be an integer $j_1$, with $1 < j_1 <kq+m-1$, such that $x_{j_1} = 1$. Then we have two words occurring in $\mathbf{x}$ of the form $1\mathbf{a}1$ and $1\mathbf{a}^\prime 1$. Assume for the moment that the two resulting words are  such that $|1\mathbf{a}1| = kq -1$ and $|1\mathbf{a}^\prime 1| = kq^\prime -1$ for some $k,k^\prime\in\mathbb{N}_0$ then, the word $1\mathbf{a}_01 = 1\mathbf{a}1\mathbf{a}^\prime 1$ would have a size $|1\mathbf{a}1| = \tilde k q -1$ with $\tilde k = k+k^\prime$ which clearly contradict the hypothesis. Then, at least one of these words should have the form $1\mathbf{a}_11$ with $ |\mathbf{a}_1| = k_1q+m_1-1$ for $1\leq m_1 \leq q-1$. By the same argument as above, it is not possible that $\mathbf{a}_1 = 0^{k_1q+m_1-1}$ and thus, there is a $j_2$, with $1 < j_2< kq+m−1$, such that $x_{j_2} = 1$. Then we use again the same argument as above, there is a word of the form $1\mathbf{a}_21$ occurring in $\mathbf{x}$. Repeating this argument successively we obtain a sequence of words $\mathbf{a}_0, \mathbf{a}_1, \mathbf{a}_2, \dots$ such that  $|\mathbf{a}_0|> |\mathbf{a}_1| >|\mathbf{a}_2| > \cdots$. Since all these words have a size of the form $kq+m-$ for some $k\in \mathbb{N}_0$ and some $1\leq m \leq q-1$, then, the smallest word $\mathbf{a}_N$ we can obtain from this procedure  is  $| \mathbf{a}_N |= 0$.  This means that the word $\mathbf{a}_N $ is the empty symbol, and therefore the word $1\mathbf{a}_N1 = 11$ occurs in $\mathbf{x}$. This is impossible because the word $11$ is  forbidden in the set $Y^{(q)}_n$ (take $k = 0$ and $m=1$ in $10^{kq+m-1}1$).

\hfill\ensuremath{\square}
%%%%%%%%%%%%%%%%%%%%%%%%%%%%%%%%%%%%%

We will prove Theorem~\ref{teo1} in two steps. First, we will show that  the set $Y^{(q)}$ is indeed the groundstate for $\phi_{\mathrm{c}}$. Then, we will prove that this set turns out to be a sofic subshift. 

%%%%%%%%%%%%%%%%%%%%%%%%%%%%%%%%%%%%%
\emph{Proof of Theorem~\ref{teo1}:} 
First let us prove that $Y^{(q)} \subset \underline{X}\left(\phi_{\mathrm{c}}\right)$. This is equivalent to prove that  ${Y}_{n}^{(q)} \subset \underline{X}\left(\phi_{\mathrm{c}}\right) $  for all $n\in\mathbb{N}$  since the closure of the union of all the $Y_n^{(q)}$ should also be a subset of $\underline{X}\left(\phi_{\mathrm{c}}\right)$ because the latter is closed. 

First let us prove that $Y_{n}^{(q)} \subset \underline{X}\left(\phi_{\mathrm{c}}\right) $ when $n$ is not a multiple of $q$. In this case, as stated in the above lemma,  $Y^{(q)}_n$ has only one element $Y^{(q)}_n = \{0^\infty\}$. But $0^\infty$ is a configuration that trivially minimize the energy, since $\phi_{\mathrm{c}}(0^\infty) = 0$.

Now let us consider the case $n = {k}q$, for some ${k}\in\mathbb{N}$, and take an element $\mathbf{x} \in Y_n^{(q)} $. We need to prove that $S_n\phi_{\mathrm{c}}(\mathbf{x}) =0$ in order for $\mathbf{x}$ to belong to the groundstate. Then, assume that $S_n\phi_{\mathrm{c}}(\mathbf{x}) > 0$. If this were the case we would have that
\[
S_n\phi_{\mathrm{c}}(\mathbf{x}) = \sum_{j=0}^{n-1}\phi_{\mathrm{c}} \circ T^j(\mathbf{x})>0.
\]
Recall that the function $\phi_{\mathrm{c}}  $ is non-negative, i.e., $\phi_{\mathrm{c}}(\mathbf{x}) \geq 0$  for all $\mathbf{x}\in X$. Thus, we can say that there is a $j_0\in\mathbb{Z}$ such that $\phi_{\mathrm{c}}\circ T^{j_0} (\mathbf{x}) >0$. Then,
\[
\phi_{\mathrm{c}} \circ T^{j_0} (\mathbf{x}) = \sum_{i=1}^\infty J(i)x_{j_0} x_{j_0+i} > 0.
\]
From the above we necessarily have that $x_{j_0} =1$, if not, we would have that $\phi_{\mathrm{c}} \circ T^{j_0} (\mathbf{x}) = 0$. Moreover, we necessarily have that there is a $i^\prime\in\mathbb{N}$ such that $x_{{j_0}+i^\prime} = 1$, in order to satisfy the above inequality. This integer $i^\prime$ cannot be chosen as a multiple of $q$, because if it were the case we would have that  $\phi_{\mathrm{c}} \circ T^{j_0} (\mathbf{x}) = 0$ since $J(pq) =0$ for all $p\in\mathbb{N}$ by definition. Then, such an integer can be written as $i^\prime= k_0q+m_0$ for some $1\leq m_0\leq q-1$ and some $k_0\in\mathbb{N}_0$. This means that a word of the form $1\mathbf{a}_01$ occurs in $\mathbf{x}$ with $|\mathbf{a}_0| = k_0q+m_0-1$. Specifically we have that  $\mathbf{x}_{j_0}^{j_0+k_0q + m_0} = 1\mathbf{a}_01$. However, the word $\mathbf{a}_0$ cannot be $0^{k_0 + m_0-1}$ because it is forbidden by hypothesis (just recall that $\mathbf{x} \in Y^{(q)}_n$). Then we follow a similar argument drawn in the proof of Lemma~\ref{lemma-1}. Since $\mathbf{a}_0$ cannot be $0^{k_0 + m_0-1}$ then there is a $j_1$ such that $(\mathbf{a}_0)_{j_1} =1$, which implies that a word of the form $1\mathbf{a}_11$ occurs in $\mathbf{x}$ with $|\mathbf{a}_1| = k_0q+m_0-1$ for some $1\leq m_1 \leq q-1$ and some $k_1 \in \mathbb{N}_0$. Repeating this procedure successively, we obtain a sequence of words $ \mathbf{a}_0, \mathbf{a}_1, \mathbf{a}_2, \dots $ with decreasing sizes, $|\mathbf{a}_0|> |\mathbf{a}_1| >|\mathbf{a}_2| > \cdots$.  Then, within this sequence there occur a word $\mathbf{a}_N$ of size $0$,   which means that the word $11$ occurs in $\mathbf{x} \in Y^{(q)}_n$. However this is impossible, because from the very definition of $Y^{(q)}_n$. This proves by contradiction that necessarily $Y^{(q)} \subset \underline{X}(\phi_{\mathrm{c}})$.

Now let us prove that $\underline{X}(\phi_{\mathrm{c}}) \subset  Y^{(q)}$.  As in the preceding case, it is enough to prove that if a $n$-periodic point belongs to $ \underline{X}(\phi_{\mathrm{c}}) $ then it should belongs to $Y^{(q)}_n$. 

First let us consider the case in which $n$ is not a multiple of $q$, i.e., that $n=kq+m$ for some $k\in \mathbb{N}_0$ and some $1\leq m\leq q-1$ . We know from Lemma~\ref{lemma-1} that in this case  $Y^{(q)}_n$ has only one element. Then we need to prove that the unique $n$-periodic point belonging to $\underline{X}(\phi_{\mathrm{c}}) $ is $0^\infty$. Assume that it is not the case and that therefore there is a $\mathbf{x} \in \underline{X}(\phi_{\mathrm{c}}) $ such that $x_{j} =1$ for some $j\in\mathbb{Z}$. By periodicity of $\mathbf{x}$ we can chose $j$ to be such that $0\leq j \leq n$. Moreover, by the periodicity of $\mathbf{x}$  we also have that $x_{j+kq+m} = 1$ for some $k\in \mathbb{N}_0$ and some $1\leq m\leq q-1$. This means that a word of the form $1\mathbf{a}1$ occurs in $\mathbf{x}$ with $|\mathbf{a}| = kq+m-1$. Then notice that
\[
S_n\phi_{\mathrm{c}}(\mathbf{x}) = \sum_{i=0}^{n-1} \phi_{\mathrm{c}} \circ T^i(\mathbf{x})  \geq  \phi_{\mathrm{c}} \circ T^j(\mathbf{x}), 
\]
which in turns implies that
\[
S_n\phi_{\mathrm{c}}(\mathbf{x}) = \sum_{i=i}^{\infty} J(i) x_{j}x_{j+i}\geq  J(kq+m) >0,  
\]
by choosing $i = kq+m$. The latter inequality contradicts the hypothesis that $x\in \underline{X}(\phi_{\mathrm{c}}) $. This proves that the only  $n$-periodic point in $\underline{X}(\phi_{\mathrm{c}}) $ is $0^\infty$.

Now let us take a $n$-periodic point $\mathbf{x}\in \underline{X}(\phi_{\mathrm{c}}) $ with $n = \tilde{k}q$ for some $\tilde{k}\in \mathbb{N}$. Assume that $\mathbf{x}$ does not belong to $Y^{(q)}_n$. The latter means that a word of the form $10^{kq+l-1}1$ occurs in $\mathbf{x}$ for some $k\in\mathbb{N}_0$ and some $1\leq l \leq q-1$. We do not loss generality if we assume that $\mathbf{x}^{kq+l}_0=10^{kq+l-1}1$ (we can apply the shift mapping successively until $10^{kq+l-1}$ be a prefix of the shifted sequence and then redefine $\mathbf{x}$). Then note that
\[
S_n\phi(\mathbf{x}) = \sum_{j=0}^{n-1}  \phi\circ T^j (\mathbf{x}) > \phi(\mathbf{x})  = \sum_{i=1}^\infty J(i) x_0 x_i >J(kq+l)>0.
\]
However, the above inequality contradicts the hypothesis which says that $\mathbf{x} \in \underline{X}(\phi_{\mathrm{c}})  $ and therefore $S_n\phi(\mathbf{x})$ should be strictly zero.  This completes the proof.
\hfill\ensuremath{\square}
%%%%%%%%%%%%%%%%%%%%%%%%%%%%%%%%%%%%%

\begin{remark}
\label{remark1}
We should observe that in the proof given above for Theorem~\ref{teo1} we obtained an additional result concerning the set of forbidden words. Indeed, to prove that a word of the form $10^{kq+m-1}1$ does not appear in the groundstate we needed to prove that, actually, a word of the form $1\mathbf{a}1$ is forbidden for any $\mathbf{a}\in X$ with $|\mathbf{a}| = kq+m-1$. Thus, the shift space $Y^{(q)}$ can be expressed either, in terms of the set $F(Y^{(q)})$ of forbidden words defined as
\[
F(Y^{(q)}) := \{ 10^{kq+m}1 \in X\, : \, k\in \mathbb{N}_0 \,\,\mbox{and}\,\, 1\leq m \leq q-1\}, 
\]
or in terms of the set $\tilde{F}(Y^{(q)})$ of forbidden words defined as,
\[
F(Y^{(q)}_n) = \{ 1\mathbf{a}1 \, : \, \forall \, \mathbf{a} \in\mathcal{A}^{kq+m-1},\, \forall \, k\in \mathbb{N}_0, \forall\, 1\leq m \leq q-1\}.
\]

\end{remark}

Now we proceed to prove Theorem~\ref{teo:t2}. We first establish the follower sets for all the admitted words of size $q$. Next, we prove that the follower sets of any other admitted word in $Y^{(q)}_n$ must be necessarily one of the follower sets of the admitted words of size $q$. 

%%%%%%%%%%%%%%%%%%%%%%%%%%%%%%%%%%%%%
\emph{Proof of Theorem~\ref{teo:t2}:} 
We now that the groundstate for $\phi_{\mathrm{c}}$ is the subshift $Y^{(q)}$ defined above. Notice that there are exactly $q+1$ admitted words of size $q$. This is easily seen because the forbidden words up to of size $q$ are $11, 101, \dots, 10^{q-2}1$. Then, it follows that admitted words of size $q$ cannot contain two one's. This implies that the set of admitted words of size $q$ is,
\[
\mathcal{B}_{q}(Y^{(q)}) = \{0^q\}\cup \{ 0^{q-m-1}10^m \, : \, 0\leq m \leq q-1\}.
\]

First we will prove that the follower sets corresponding to these words are given by Eq.~(\ref{eq:followersets}). Let us start with the follower sets of $0^q$. Let $n\in \mathbb{N}$, then it is easy to see that any the admitted word $\mathbf{a}$ of size $n$ in ${Y^{(q)}}$ are such that $0^q\mathbf{a}$ is also an admitted word. This is because the concatenation  $0^q\mathbf{a}$ does not generates a forbidden words, because forbidden words $Y^{(q)}$ are all of the form $10^{kq+m}1$ for all $k\in \mathbb{N}_0$ and all $1\leq m \leq q-1$ (there are no ``one's'' in $0^q$ giving rise to forbidden word after the concatenation). This implies that $ \mathcal{B}_n(Y^{(q)}) \subset  \mathcal{F}(0^q) $ for all $n\in \mathbb{N}$. The latter immediately implies that $  \mathcal{F}(0^q)  = \mathcal{B}(Y^{(q)}) $ since by definition every follower set is a subset of $\mathcal{B}(Y^{(q)})$. 

Now we will prove that 
\begin{eqnarray}
\mathcal{F}(0^{m}10^{q-m-1}) = \bigcup_{n\in\mathbb{N}} \mathcal{C}_{n,m}^{(q)},
\nonumber
\end{eqnarray}
where the sets  $\mathcal{C}_{n,m}^{(q)}$ are given by,
\[
\fl
\mathcal{C}_{n,m}^{(q)} := \left\{ \mathbf{a} = a_0a_1a_2\dots a_{n-1} \in \mathcal{B}_n(Y^{(q)}) \, : \, a_i =0 \, \forall \,i \not= kq + m,\, \forall\, 0\leq k\leq \lfloor n/q \rfloor\right\}.
\]
First let us show that $\mathcal{C}_{n,m}^{(q)} \subset \mathcal{F}(0^{m}10^{q-m-1})  $ for every $n\in\mathbb{N}$ and for every $0\leq m \leq q-1$. Take an element $\mathbf{a}\in \mathcal{C}_{n,m}^{(q)}$ and assume that the word $0^{m}10^{q-m-1} \mathbf{a}$ give rise some forbidden word. Since $\mathbf{a}$ does not contain forbidden words, the only possibility is that the forbidden word arises due to the presence of the `1' in $0^{m}10^{q-m-1}$. This means that the forbidden word should be of the form $\mathbf{w} = 10^{q-m-1}a_0a_1a_2 \dots a_{j-1} a_j $ for some $1\leq j \leq n$. If this is the case, then it is clear that $a_j = 1$ and $a_i = 0$ for all $0\leq i \leq {j-1}$. Thus, the word $\mathbf{w}$ takes the form,
\[
\mathbf{w} = 10^{q-m-1}a_0a_1a_2 \dots a_{j-1} a_j = \left( 10^{q-m-1 }\right) \left( 0^{j } 1\right) = 10^{ q-m + j -1}1.
\]
Since $\mathbf{w}$ is forbidden, it is clear that $ q-m+j-1 $ should be of the form $ k^\prime q +l - 1$, for some $k^\prime\in \mathbb{N}_0$ and some $1\leq l\leq q-1$. Next have that  $q-m+j-1 = k^\prime q + l -1 $ implies that 
\begin{equation}
j = (k^\prime-1) q + m + l .
\label{eq:j1}
\end{equation}
On the other hand, since $\mathbf{a}$ belongs to $\mathcal{C}_{n,m}^{(q)}$  hence we have that $a_{i} = 0 $  $\forall \,i \not= kq + m,\, \forall\, 0\leq k\leq \lfloor n/q \rfloor$. Due to the fact that $a_j = 1$, we have that there is a $ 0\leq k\leq \lfloor n/q \rfloor$ such that  $j = kq + m $. However, in view of Eq.~(\ref{eq:j1}) we see that $l$ should be zero in order to fulfill both equations for $j$. But $l=0$ contradicts the hypothesis that $\mathbf{w} =  10^{ q-m + j -1}1 = 10^{k^\prime q +l - 1}1$ is a forbidden word. This proves that $\mathcal{C}_{n,m}^{(q)} \subset \mathcal{F}(0^{m}10^{q-m-1}) $. 

Now let us prove the reverse inclusion, $\mathcal{F}(0^{m}10^{q-m-1})  \subset \mathcal{C}_{n,m}^{(q)} $.  Take an element $\mathbf{a} \in \mathcal{F}(0^{m}10^{q-m-1}) $ and assume that it is not contained in $ \mathcal{C}_{n,m}^{(q)} $. Since $\mathbf{a} \not\in \mathcal{C}_{n,m}^{(q)} $, we have that $a_i = 1$ for some $i \not= kq + m$ with $0\leq k\leq \lfloor n/q \rfloor $. But the concatenated word $0^{m}10^{q-m-1} \mathbf{a}$ has a word $\mathbf{w}$ of the form
\[
10^{q-m-1} a_1a_2\dots a_{i-1} a_i = 10^{q-m-1} a_0 a_1a_2\dots a_{i-1}1. 
\]
Because the size of the word $0^{q-m-1}a_0 a_1a_2\dots a_{i-1} $ is $ q- m + i - 1$ we have that, in order for $\mathbf{w}$ be admitted, we require that $ q- m + i - 1 = \tilde {k} q  -1$ for some $\tilde k\in \mathbb{N}$. The latter implies that $ i = (\tilde  k-1) q + m $. This is not possible by hypothesis, since we assumed that  $\mathbf{a} \not\in \mathcal{C}_{n,m}^{(q)} $, and hence $i \not= kq + m$ for any $0\leq k\leq \lfloor n/q \rfloor $.

Next we prove that given any other word in the language $\mathcal{B}(Y^{(q)})$ has a follower set that coincides with one of the already given for the admitted words of size $q$.

Let $\mathbf{a} = a_0 a_1 a_2 \dots a_n$ be an admitted word, i.e., $\mathbf{a} \in \mathcal{B}(Y^{(q)})$.  We need to prove that the follower set $\mathcal{F}(\mathbf{a})$ is one of the follower sets already described for the words of size $q$. If the word $\mathbf{a}$ turns out to be $0^{n+1}$, it is clear that all the words in the language of $Y^{(q)}$ can be a follower of $\mathbf{a}$. This means that $\mathcal{F}(0^{n})=\mathcal{F}(0^q)=\mathcal{B}(Y^{(q)})$. 

Now, let us consider the case in which there is a ${j}^*$ such that $a_{n-{j}^*}=1$ for  some $0\leq {j}^*\leq n$. Without loss of generality we can write ${j^*}={k^*}q-{m^*}-1$ for some $0\leq {m^*} \leq q-1$ and some ${k^*}\in\mathbb{N}_0$. We will prove that the follower set for $\mathbf{a}$, $\mathcal{F}(\mathbf{a})$  coincides with $\mathcal{F}(0^{{m^*}}10^{q-{m^*}-1})$. To prove the latter assume that the word $\mathbf{ab}$ contains a forbidden word for some $\mathbf{b}\in\mathcal{F}(0^{{m^*}}10^{q-{m^*}-1})$. We can assume without loss of generality that the $(n-{j^*})$th coordinate of $\mathbf{a}$ and the ${i^*}$th coordinate of $\mathbf{b}$, are the symbols that generate the forbidden word in the concatenation $\mathbf{ab}$, i.e,
\[
\mathbf{a}_{{n-j^*}}^{n}\mathbf{b}_0^{{i^*}} = 1\mathbf{w}1,
\]
where,
\[
\mathbf{w} :=a_{{n-j^*}+1}a_{{n-j^*}+2}\dots a_n b_0 b_1 \dots b_{{i^*}-1}.
\]
Since $1\mathbf{w}1$ is forbidden, then, according to Remark~\ref{remark1}, we have that the size of $\mathbf{w}$, which is given by $|\mathbf{w}| = j^* +i^* $, should be of the form $kq+m-1$ for some $k\in\mathbb{N}_0$ and some $1\leq m \leq q-1$. This means that $j^* +i^* = {k^*}q-{m^*}-1 + i^* = kq+m-1$, condition that can be rewritten as,
\begin{equation}
\label{condition-i-1}
i^* = (k-k^*)q+m+m^*.
\end{equation}
Now, recall that $\mathbf{b}$ is in  $\mathcal{F}(0^{{m^*}}10^{q-{m^*}-1})$, which means that $b_i =0 $ for all $i\not={k}q+{m^*}$ for all $0\leq {k}\leq \lfloor n/q\rfloor$. Since $b_{i^*} =1$, this means that $i^*$ should be of the form $i^* = \tilde{k} q + {m^*}$ for some $0\leq \tilde{k} \leq \lfloor n/q\rfloor$. This last condition together with the requirement~(\ref{condition-i-1}) implies that $m=0$, which is not possible by hypothesis. This implies that $\mathbf{b}$ is necessarily a follower of $\mathbf{a}$, which shows that $ \mathcal{F}(0^{{m^*}}10^{q-{m^*}-1}) \subset \mathcal{F}(\mathbf{a})$. 

Now, take an element $\mathbf{c} \in \mathcal{F}(\mathbf{a})$. We will prove that  $\mathbf{c} \in \mathcal{F}(0^{{m^*}}10^{q-{m^*}-1})$ if, as above, $\mathbf{a}$ is such that $a_{n-{j}^*}=1$ for  some $0\leq {j}^*\leq n$, with ${j^*}={k^*}q-{m^*}-1$ for some $0\leq {m^*} \leq q-1$ and some ${k^*}\in\mathbb{N}_0$. If there is another coordinate of $\mathbf{a}$ is such that $a_{n-j^\prime} =1$, for some $j^\prime > j^*$ then notice that the word $1\mathbf{a}_{n-j^\prime+1}^{n-j^*-1}1$  is not forbidden, which means that $|\mathbf{a}_{n-j^\prime+1}^{n-j*-1}1| = j^\prime -j^* -1 $ should be necessarily of the form $k^\prime q - 1$. This implies that $j^\prime - j^*-1 = j^\prime - {k^*}q +{m^*} = k^\prime q - 1$, or equivalently 
\[
j^\prime = (k^\prime-k^*)q -m-1.
\]
A similar result is obtained under the contrary assumption, $j^\prime < j^*$. This result that the choice of $m^*$ is unique, independently of which coordinate is equal to one along $\mathbf{a}$. 

Since $\mathbf{c}$ does not belong to $\mathcal{F}(0^{{m^*}}10^{q-{m^*}-1})$ by hypothesis, we have that $c_{kq + m^* +l} = 1$ for some $k\in \mathbb{N}_0$ and some $1\leq l \leq q-1$. But the word  $\mathbf{c}$ is a follower of $\mathbf{a}$, then $\mathbf{ac}$ does not contain any forbidden word. Since $a_{n-j*}=1$ and $c_{kq + m^* +l} = 1$ we need to verify that the word
\[
\mathbf{a}_{n-j^*}^{n}\mathbf{c}_{0}^{kq + m^* +l} = 1\mathbf{w}1
\]
is not forbidden. In this case we have defined $\mathbf{w}$ as $\mathbf{a}_{n-j^*+1}^{n}\mathbf{c}_{0}^{{kq + m^* +l}-1}$. In order for $1\mathbf{w}1$ to be admitted we need at least that $|\mathbf{w}| = {kq + m^* +l}+j^*$ be of the form $\tilde{k}q -1$ for some $\tilde{k}\in\mathbb{N}_0$. This condition is equivalent to 
\[
{kq + m^* +l}+j^* =  (k-k^*)q +l - 1 = \tilde{k}q -1.
\]
The latter implies that $l=0$ or $l={k^\prime} q$ for some $k\in\mathbb{N}$. But this is impossible, since we assumed that $1\leq l \leq q-1$. This proves that $\mathbf{c}$ belongs to  $\mathcal{F}(0^{{m^*}}10^{q-{m^*}-1})$. 

\hfill\ensuremath{\square}
%%%%%%%%%%%%%%%%%%%%%%%%%%%%%%%%%%%%%

\begin{figure}[ht]
\begin{center}
\scalebox{0.3}{\includegraphics{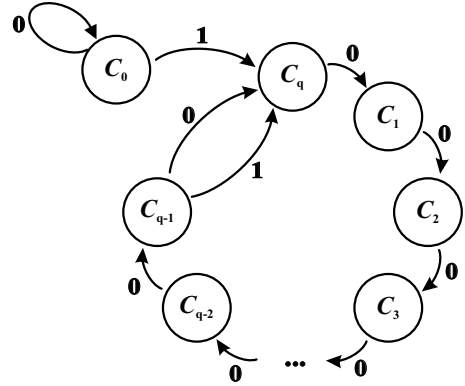}}
\end{center}
     \caption{
     Right-resolving presentation for the sofic subshift $Y^{(q)}$, which corresponds to the groundstate for the potential $\phi_\mathrm{c}$.
             }
\label{fig:presentationYq}
\end{figure}

Finally we calculate the topological entropy as follow. It is known that the follower sets of a given sofic subshift allows us to build up a right-resolving presentation. Actually, there is a standard way to do this~\cite{lind1995introduction}. The vertices of the graph are represented by the follower sets given in Theorem~\ref{teo:t2}. To label the vertex of the graph we will use the following short-hand notation,
\begin{eqnarray}
C_m &:=& \mathcal{F}(0^{q-m}10^m-1) \qquad \mbox{for } \qquad 1\leq m \leq q,
\nonumber
\\
C_0 &:=& \mathcal{F}(0^{q}).
\end{eqnarray}
Now, take a vertex, say for example $C_m = \mathcal{F}(\mathbf{a})$, where $\mathbf{a}$ is one of the admitted words of size $q$, and determine the follower sets of the words $\mathbf{a}1$ and $\mathbf{a}0$. The follower sets $\mathcal{F}(\mathbf{a}1)$ and $\mathcal{F}(\mathbf{a}0)$ are empty or belong necessarily to the family  $\{C_i : 0\leq i \leq q\}$. Assume for the moment that such follower sets are not empty. Then there are $j,k$ such that $C_{j} =\mathcal{F}(\mathbf{a}0) $ and $C_{k}:= \mathcal{F}(\mathbf{a}1)$. Then draw an edge from $C_m$ to $C_j$ labeled with `0' and draw an edge from $C_m$ to $C_k$ labeled with `1'. On the other hand if either $\mathcal{F}(\mathbf{a}1)$ or $\mathcal{F}(\mathbf{a}0)$ is empty,  then no edge corresponding to such label is drawn. Repeat this procedure for all the vertices.  With this procedure we obtain the \emph{follower set graph} (an edge labeled graph) $\mathcal{G}_q$ that turns out to be a right-resolving presentation~\cite{lind1995introduction} for sofic subshift $Y^{(q)}$. This graph is shown in Fig.~\ref{fig:presentationYq}. The follower set graph allows us to write down the adjacency matrix $A_q$ corresponding to this labelled graph. The adjacency matrix for $\mathcal{G}_q$ is given by

\begin{equation}
A_q = \left[
  \begin{array}{cccccccc}
   1 & 0 & 0& 0 & \ldots & 0& 0 & 1 \\
   0 & 0 & 1 & 0 & \ldots & 0 & 0 & 0 \\
   0 & 0 & 0 & 1 & \ldots  & 0 & 0 & 0 \\
   \vdots & \vdots & \vdots& \vdots &\ddots & \vdots &\vdots &\vdots  \\
   0 & 0 & 0 & 0 & \ldots & 0 & 1 & 0 \\
   0 & 0 & 0 & 0 & \ldots & 0 & 0 & 2 \\
   0 & 1 & 0 & 0 & \ldots & 0 &0 & 0 \\
  \end{array} \right]
\end{equation}

%%==================== FIGURE =========================
%%
%%==================== FIGURE =========================
%

It is not difficult to obtain the eigenvalues of $A_q$. After some calculations we can observe that the characteristic polynomial determining the eigenvalues can be written as,
\[
(1-\lambda) (2-\lambda^{q})=0.
\]
The largest eigenvalue $\lambda_0$ of $A_{q}$ gives the topological entropy as follows (see Ref.~\cite{lind1995introduction}),
\[
h_{\mathrm{top}} = \log(\lambda_0) = \log(2^{1/q}) = \frac{\log(2)}{q},
\]
from which it follows immediately the result stated in Corollary~\ref{coro-1}.

\bigskip

%#############################################################
%#############################################################
%
\section*{Acknowledgements} 
%
%#############################################################
%#############################################################

The authors thanks CONACyT for financial support through Grant No. CB-2012-01-183358.

\section*{References}
\nocite{*}
\bibliography{DegeneratedGS_ref.bib}% Produces the bibliography via BibTeX.

%
%\section*{References}
%\begin{thebibliography}{10}
%
%\bibitem{book1} Goosens M, Rahtz S and Mittelbach F 1997 {\it The \LaTeX\ Graphics Companion\/} 
%(Reading, MA: Addison-Wesley)
%
%\bibitem{eps} Reckdahl K 1997 {\it Using Imported Graphics in \LaTeX\ } (search CTAN for the file `epslatex.pdf')
%
%\end{thebibliography}
%
%

\end{document}